\RequirePackage{lineno}
\documentclass[aps,prd,twocolumn,superscriptaddress,showpacs,amsmath,amssymb]{revtex4-1}

\usepackage{amsmath,amssymb}
\usepackage{comment}
\usepackage{multirow}
\usepackage{dcolumn}% Align table columns on decimal point
\usepackage{bm}% bold math
\usepackage{setspace}
\usepackage{graphicx}% Include figure files
\includecomment{printrobustnotes}
\includecomment{printallnotes}
\usepackage{xspace}

\usepackage{dcolumn}% Align table columns on decimal point
\usepackage{bm}% bold math
\newenvironment{cfigure1c}[1][tbp]{\begin{figure*}[#1]\centering}{\end{figure*}}

\newcommand{\note}[1]{\xspace}
\newcommand{\robustnote}[1]{\xspace}
\begin{printrobustnotes}
\renewcommand{\robustnote}[1]{\textbf{\textit{#1}}\xspace}
\end{printrobustnotes}
\begin{printallnotes}
\renewcommand{\note}[1]{\textbf{\textit{#1}}\xspace}
\renewcommand{\robustnote}[1]{\textbf{\textit{#1}}\xspace}
\end{printallnotes}
\newcommand{\gevcc}[1]  {\ensuremath{#1~\mathrm{GeV}/c^{2}}}

\newcommand{\gevccnoarg}{\ensuremath{\mathrm{GeV}/c^{2}}\xspace}

\begin{document}
%\setpagewiselinenumbers
%\modulolinenumbers[1]
%\linenumbers

%\begin{frontmatter}

\title{Search for low-mass dark matter with CsI(Tl) crystal detectors}

%\author{KIMS Collaboration}
\affiliation{Department of Physics, Ewha Womans University, Seoul 120-750, Korea}
\affiliation{Department of Physics and Astronomy, Seoul National University, Seoul 151-747, Korea}
\affiliation{Department of Science Education, Ewha Womans University, Seoul 120-750, Korea}
\affiliation{Center for Underground Physics, Institute for Basic Science (IBS), Daejon 305-811, Korea}
\affiliation{Department of Physics, Kyungpook National University, Daegu 702-701, Korea}
\affiliation{Department of Physics, Sejong University, Seoul 143-747, Korea}
\affiliation{Korea Research Institute of Standards and Science, Daejon 205-340, Korea}
\affiliation{Department of Physics, The University of Seoul, Seoul 130-743, Korea}
\affiliation{Department of Engineering Physics, Tsinghua University, Beijing 100084, China}
\affiliation{Institute of High Energy Physics (IHEP), Beijing 100049, China}

\author{H.S.~Lee}
\email[Corresponding Author : ]{hyunsulee@ewha.ac.kr}
\affiliation{Department of Physics, Ewha Womans University, Seoul 120-750, Korea} 
\author{H.~Bhang}
\affiliation{Department of Physics and Astronomy, Seoul National University, Seoul 151-747, Korea} 
\author{J.H.~Choi}
\affiliation{Department of Physics and Astronomy, Seoul National University, Seoul 151-747, Korea}
\author{S.~Choi}
\affiliation{Department of Physics and Astronomy, Seoul National University, Seoul 151-747, Korea} 
\author{I.S.~Hahn}
\affiliation{Department of Science Education, Ewha Womans University, Seoul 120-750, Korea} 
\author{E.J.~Jeon}
\affiliation{Center for Underground Physics, Institute for Basic Science (IBS), Daejon 305-811, Korea}
\author{H.W.~Joo}
\affiliation{Department of Physics and Astronomy, Seoul National University, Seoul 151-747, Korea}
\author{W.G.~Kang}
\affiliation{Center for Underground Physics, Institute for Basic Science (IBS), Daejon 305-811, Korea}
\author{B.H.~Kim}
\affiliation{Department of Physics and Astronomy, Seoul National University, Seoul 151-747, Korea}
\author{G.B.~Kim}
\affiliation{Department of Physics and Astronomy, Seoul National University, Seoul 151-747, Korea}
\author{H.J.~Kim}
\affiliation{Department of Physics, Kyungpook National University, Daegu 702-701, Korea}
\author{J.H.~Kim}
\affiliation{Department of Physics and Astronomy, Seoul National University, Seoul 151-747, Korea}
\author{K.W.~Kim}
\affiliation{Department of Physics and Astronomy, Seoul National University, Seoul 151-747, Korea}
\author{S.C.~Kim}
\affiliation{Department of Physics and Astronomy, Seoul National University, Seoul 151-747, Korea}
\author{S.K.~Kim}
\affiliation{Department of Physics and Astronomy, Seoul National University, Seoul 151-747, Korea}
\author{Y.D.~Kim}
\affiliation{Center for Underground Physics, Institute for Basic Science (IBS), Daejon 305-811, Korea}
\affiliation{Department of Physics, Sejong University, Seoul 143-747, Korea}
\author{Y.H.~Kim}
\affiliation{Center for Underground Physics, Institute for Basic Science (IBS), Daejon 305-811, Korea}
\affiliation{Korea Research Institute of Standards and Science, Daejon 205-340, Korea}
\author{J.H.~Lee}
\affiliation{Department of Physics and Astronomy, Seoul National University, Seoul 151-747, Korea}
\author{J.K.~Lee}
\affiliation{Department of Physics and Astronomy, Seoul National University, Seoul 151-747, Korea}
\author{S.J.~Lee}
\affiliation{Department of Physics and Astronomy, Seoul National University, Seoul 151-747, Korea}
\author{D.S.~Leonard}
\affiliation{Department of Physics, The University of Seoul, Seoul 130-743, Korea}
\author{J.~Li}
\affiliation{Center for Underground Physics, Institute for Basic Science (IBS), Daejon 305-811, Korea}
\author{J.~Li}
\affiliation{Department of Engineering Physics, Tsinghua University, Beijing 100084, China}
\author{Y.J.~Li}
\affiliation{Department of Engineering Physics, Tsinghua University, Beijing 100084, China}
\author{X.R.~Li}
\affiliation{Institute of High Energy Physics (IHEP), Beijing 100049, China}
\author{S.S.~Myung}
\affiliation{Department of Physics and Astronomy, Seoul National University, Seoul 151-747, Korea}
\author{S.L.~Olsen}
\affiliation{Department of Physics and Astronomy, Seoul National University, Seoul 151-747, Korea}
\author{J.W.~Park}
\affiliation{Department of Physics and Astronomy, Seoul National University, Seoul 151-747, Korea}
\author{I.S.~Seong}
\affiliation{Department of Physics and Astronomy, Seoul National University, Seoul 151-747, Korea}
\author{J.H.~So}
\affiliation{Center for Underground Physics, Institute for Basic Science (IBS), Daejon 305-811, Korea}
\author{Q.~Yue}
\affiliation{Department of Engineering Physics, Tsinghua University, Beijing 100084, China}
\collaboration{KIMS Collaboration}
\date{\today}
%\cortext[hslee]{Corresponding author. Tel:+82-2-3277-3413; fax:+82-2-3277-2372}
\begin{abstract}
We present a search for low-mass ($\leq\gevcc{20}$) weakly interacting massive particles~(WIMPs), 
strong candidates of dark matter particles,
using the low-background CsI(Tl) detector array of the Korea Invisible Mass Search experiment. 
With a total data exposure of 24,524.3~kg$\cdot$days,
we search for WIMP interaction signals produced by nuclei recoiling from WIMP-nuclear elastic scattering with visible energies between 2 and 4~keVee~(electron-equivalent energy). The observed energy distribution of candidate events is consistent with null signals, and upper limits of the WIMP-proton spin-independent interaction are set with a 90\% confidence level. The observed limit covers most of the low-mass region of parameter space favored by the DAMA annual modulation signal assuming the standard halo model. 
\end{abstract}

 % PACS, high energy physics
%\begin{keyword}
%WIMP \sep dark matter \sep CsI(Tl) Crystal \sep KIMS
%\PACS 95.35.+d\sep 14.80.Ly % PACS, high energy physics
%\end{keyword}

%\end{frontmatter}
%\PACS 95.35.+d\sep 14.80.Ly % PACS, high energy physics
%\pacs{95.35.+d,14.80.Ly} % PACS, high energy physics
\maketitle
\section{Introduction}
The existence of nonbaryonic cold dark matter has been widely supported by many astronomical observations~\cite{dm1,dm2,wmap,planck}. Theoretically favored dark matter candidates are weakly interacting massive particles~(WIMPs), which are well motivated by supersymmetric models~\cite{susydm,pdg}. In the constrained minimal supersymmetric standard model, the lightest supersymmetric particle is a WIMP candidate with an expected mass of $M_{\chi} \geq \gevcc{100}$~\cite{hWIMP}. However, there have recently been a number of experimental observations that have been interpreted as signals from WIMPs with a mass of about \gevcc{10} and a WIMP-proton spin-independent cross section of about $10^{-4}$~pb~\cite{DAMA, CRESST, CoGent, CDMS-low}. Because recent astronomical gamma-ray observations can also be interpreted as evidence for low-mass WIMPs~\cite{astro_ldm}, a low-mass WIMP as a dark matter particle candidate persists as an encouraging hypothesis~\cite{lowmass1,lowmass2,lowmass3,lowmass4,lowmass5}. Even though some experiments report null signals in this region~\cite{XENON-low,texono,LUX,CDMSlite}, 
it remains important to search for low-mass WIMPs with different types of detectors because of nontrivial systematic differences in detector responses~\cite{det1,det2} and the commonly used astronomical model for the WIMP distribution~\cite{halo}.

\section{KIMS experiment}
The Korea Invisible Mass Search (KIMS) Collaboration is performing direct searches for WIMPs using a 12-module array of low-background CsI(Tl) detectors with a total mass of 103.4~kg in the Yangyang Underground Laboratory with an earth overburden of 700~m~(2400~m water equivalent)~\cite{kims_pro}. 
The KIMS Collaboration carried out extensive research and development to identify and reduce the internal background in CsI(Tl) crystals~\cite{kims_bg,kims_nim}.
Each detector module consists of a low-background CsI(Tl) crystal with dimensions 8$\times$8$\times$30~cm$^3$ and with a green-enhanced photomultiplier tube (PMT) mounted at each end.
The crystal array is surrounded by a shield consisting, from inside out, of 10~cm of copper, 5~cm of polyethylene, 15~cm of lead, and 30~cm of liquid-scintillator-loaded mineral oil to stop external neutrons and gammas and veto cosmic-ray muons. 
Amplified signals from the PMTs of each crystal module are encoded by a 400~MHz flash analog-to-digital converter for a 32~$\mu$s time interval. Using a 59.54~keV $\gamma$ calibration from a $^{241}$Am source, we obtain a photoelectron~(PE) yield of approximately 5~PE/keVee depending on the crystal, where keVee is an electron-equivalent energy measured in the detector module. 
The trigger condition is two or more PEs in each of a crystal's two PMTs within the same 2~$\mu$s time window, corresponding to four or more PEs in the detector module, and an energy threshold of about 1~keVee.
The 12 crystals operated stably between September 2009 and December 2012. In December 2012, the operation was temporarily halted in order to upgrade the detector modules.
Details of the KIMS experiment and its WIMP search results can be found in previous publications~\cite{kims_plb,kims_prl,kims_prl2}.

The most recent KIMS experimental result, based on a partial data set with a corresponding exposure of 24,524.3~kg$\cdot$days, excluded the allowed region of parameter space that attributes the DAMA annual modulation signal to WIMPs with masses greater than \gevcc{20}~\cite{kims_prl2}.
However, these KIMS results did not establish limits for low-mass WIMPs because the analysis method utilizes pulse-shape discrimination~(PSD) of the PE time distribution to separate the nuclear recoil signals from electron recoil backgrounds~\cite{kims_psd}. In that analysis, WIMP signals are extracted using PSD parameters only, with no constraints on the background energy spectrum. 
The technique requires a minimum number of detected  PEs and, as a result, an analysis threshold that was set at 3~keVee. However, it is possible to use existing lower-energy KIMS data if the PSD requirements are  not applied.
In this paper, we present the results of a search for low-mass WIMPs using a 2~keVee energy threshold applied to data collected by the KIMS detector in 2009--2010.
This analysis uses the same data set and event-selection criteria as the recent KIMS publication~\cite{kims_prl2} but looks for low-mass WIMP signals without extracting the nuclear recoil events using PSD requirements.
To search for low-mass WIMPs, we only consider events between 2 and 4~keVee and extract potential WIMP signals from the energy distribution of the selected events.

\section{Data Analysis}

\begin{cfigure1c}
\begin{tabular}{cc}
\includegraphics[width=0.48\textwidth]{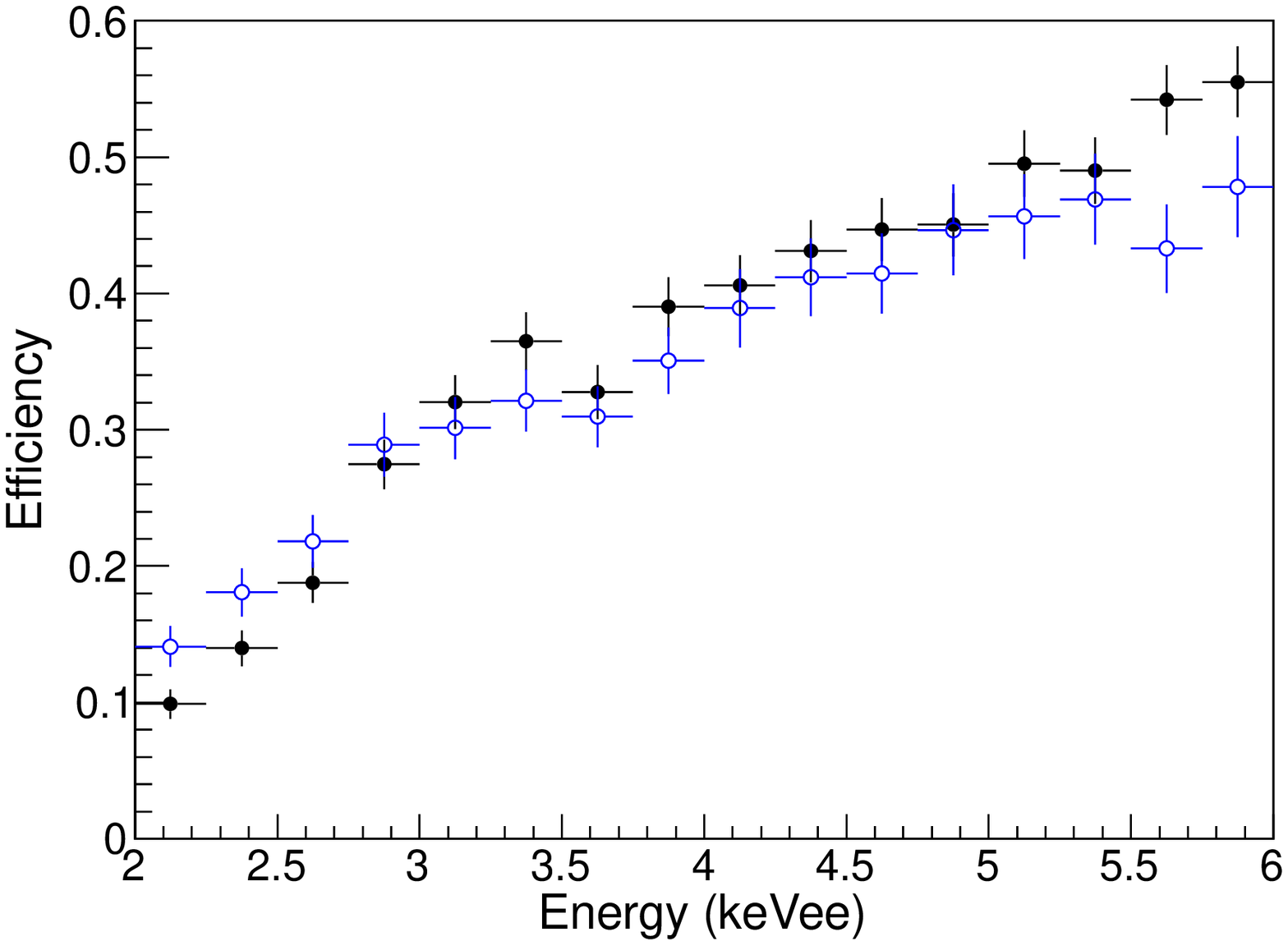}&
\includegraphics[width=0.48\textwidth]{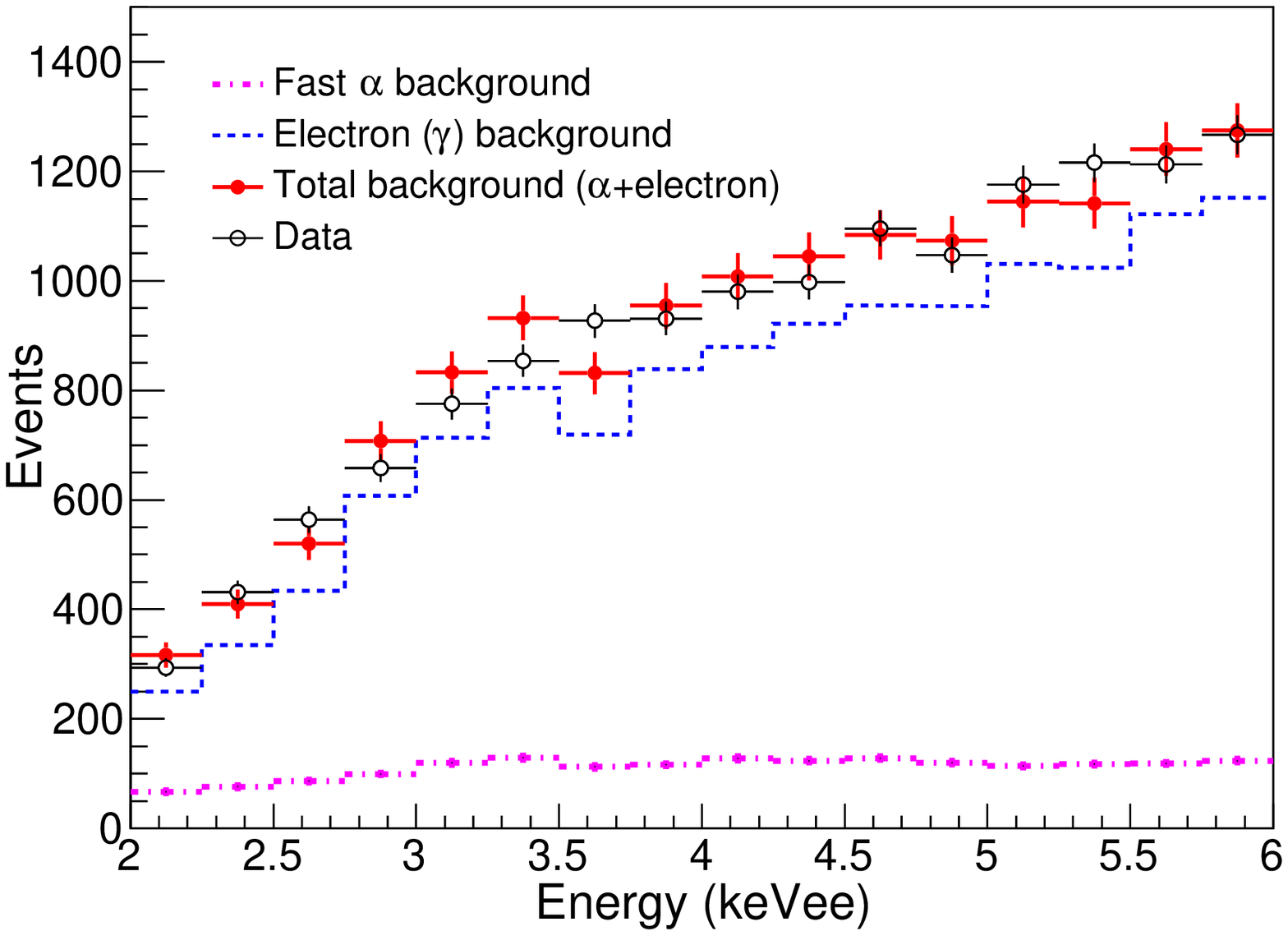}\\
(a) & (b) \\
\end{tabular}
\caption[Data fit]{
(a) The efficiencies including events trigger and selection for electron-equivalent events (filled circles) and nuclear recoil events (open circles) are presented. 
(b) The  energy distribution for selected events from one of the detector modules (open circles) is compared with the predicted background~(filled circles).
}
\label{ref:fit}
\end{cfigure1c}

%\begin{figure}
%\begin{center}
%\includegraphics[width=\columnwidth]{Edist_det6.eps}
%\end{center}
%\caption[BG shape]{
%(color online) The  energy distribution for selected events from one of the detector modules (open circles) is compared with the predicted background~(filled circles).
%}
%\label{ref:fit}
%\end{figure}

For energies below 10~keVee, PMT noise produces a significant contribution to the background. To characterize and reject PMT-noise-related background, we use a PMT ``dummy'' detector with the exact same dimension as a normal detector module but with the CsI(Tl) crystal replaced with an empty transparent acrylic box. The single dummy detector module was installed with the CsI(Tl) detector array and operated simultaneously.
We developed a set of event-selection criteria that reject the PMT-induced background signals using events that are recorded by the dummy detector~\cite{kims_prl2}. In addition, we remove cosmic-ray muon events by rejecting events that have time-associated signals in a surrounding array of liquid scintillation detectors; we  also veto events with energy deposits in more than one detector module. 
In this standard selection, all events from the PMT dummy module are rejected. %Figure~\ref{ref:fit}~(a) shows an event selection efficiency of one of the detector modules evaluated. 

After the PMT-related background is rejected with the standard selection, electron-equivalent events from $\gamma$ and $\beta$ emitters are the main sources of background. 
Because there are no known low-energy sources affecting the WIMP search data,
we modeled the energy spectrum of single-hit electron-equivalent events using multiple scattering events in which two or more detector modules satisfy the trigger condition.
Most of the single-hit events in the CsI(Tl) detector originate from Compton scattering of high-energy $\gamma$ rays and $\beta$ electrons from high Q-value $\beta$ decays~\cite{kims_nim}. 
These are expected to produce an almost flat energy spectrum in the 2--4~keVee energy region. 
This is similar for low-energy multiple-module scattering events that also originate from Compton scattering of high-energy $\gamma$ particles.
The selection efficiency for these types of events is estimated from the multiple-module scattering event spectrum shown as filled circles in Fig.~\ref{ref:fit}~(a).
%Radioactive isotopes which adhere to the crystal surfaces emit surface-$\alpha$ particles. 
Surface $\alpha$ particles that originate from radioactive isotopes that adhere to the crystal surfaces can contribute to the WIMP search data as background components. 
We characterized the surface $\alpha$ events using test crystals that were deliberately contaminated with radon progenies as described in Ref.~\cite{kims_surface}.
In order to estimate each background component, we performed a fit to the distribution of events in the 4--6~keVee energy range using PSD information as we did in the PSD analysis~\cite{kims_prl,kims_prl2}, and extrapolated the results to the 2--4~keVee region. %assuming obtained energy spectra. 
In this fit, we assume that there are two background components (electron-equivalent events and surface $\alpha$ events) with no WIMP signal. 
Figure~\ref{ref:fit}~(b) shows the energy distribution of the WIMP search data together with the prior background estimate using the no-WIMP hypothesis for one of the detector modules. 

\begin{cfigure1c}
\begin{tabular}{cc}
\includegraphics[width=0.48\textwidth]{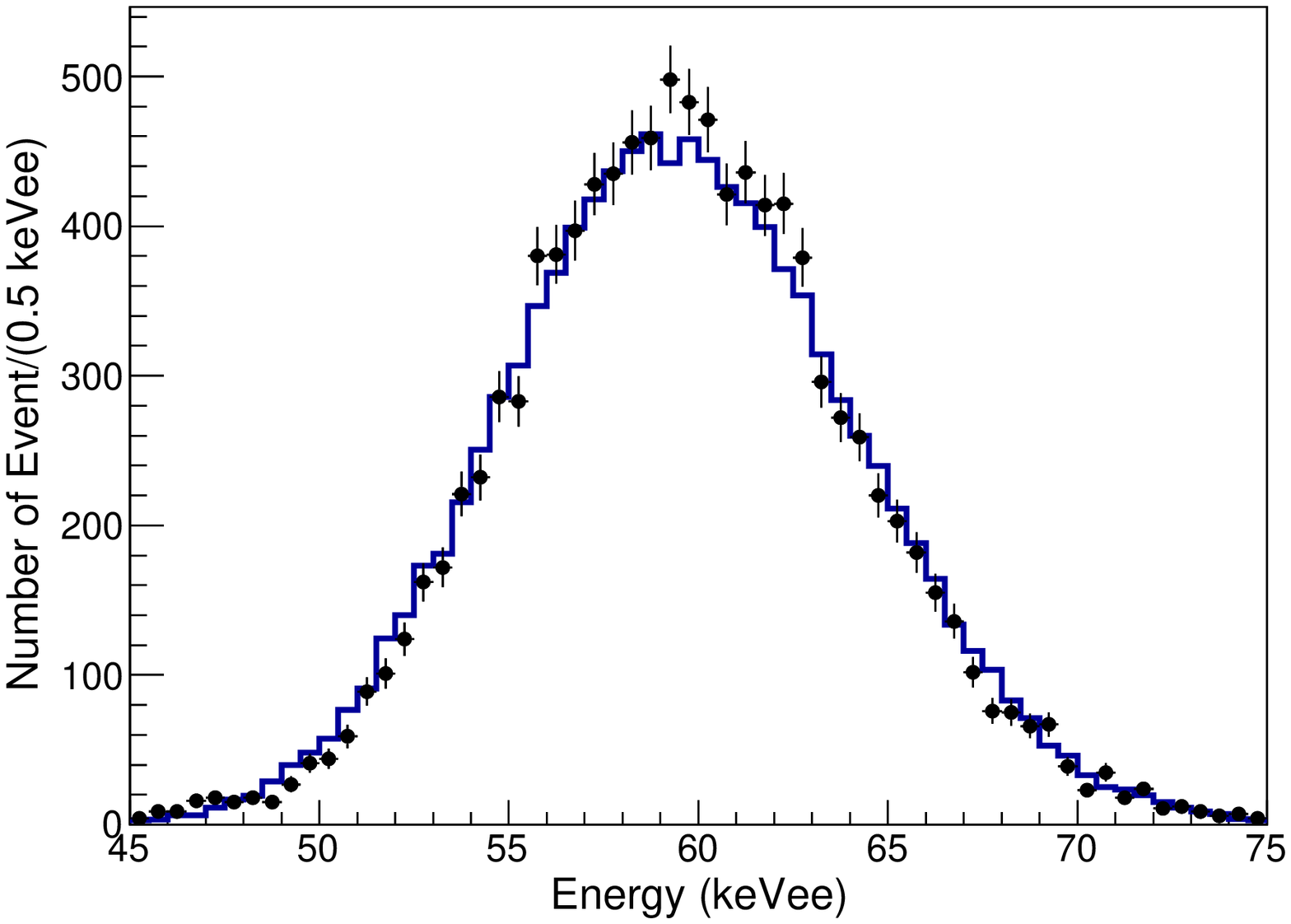}&
\includegraphics[width=0.48\textwidth]{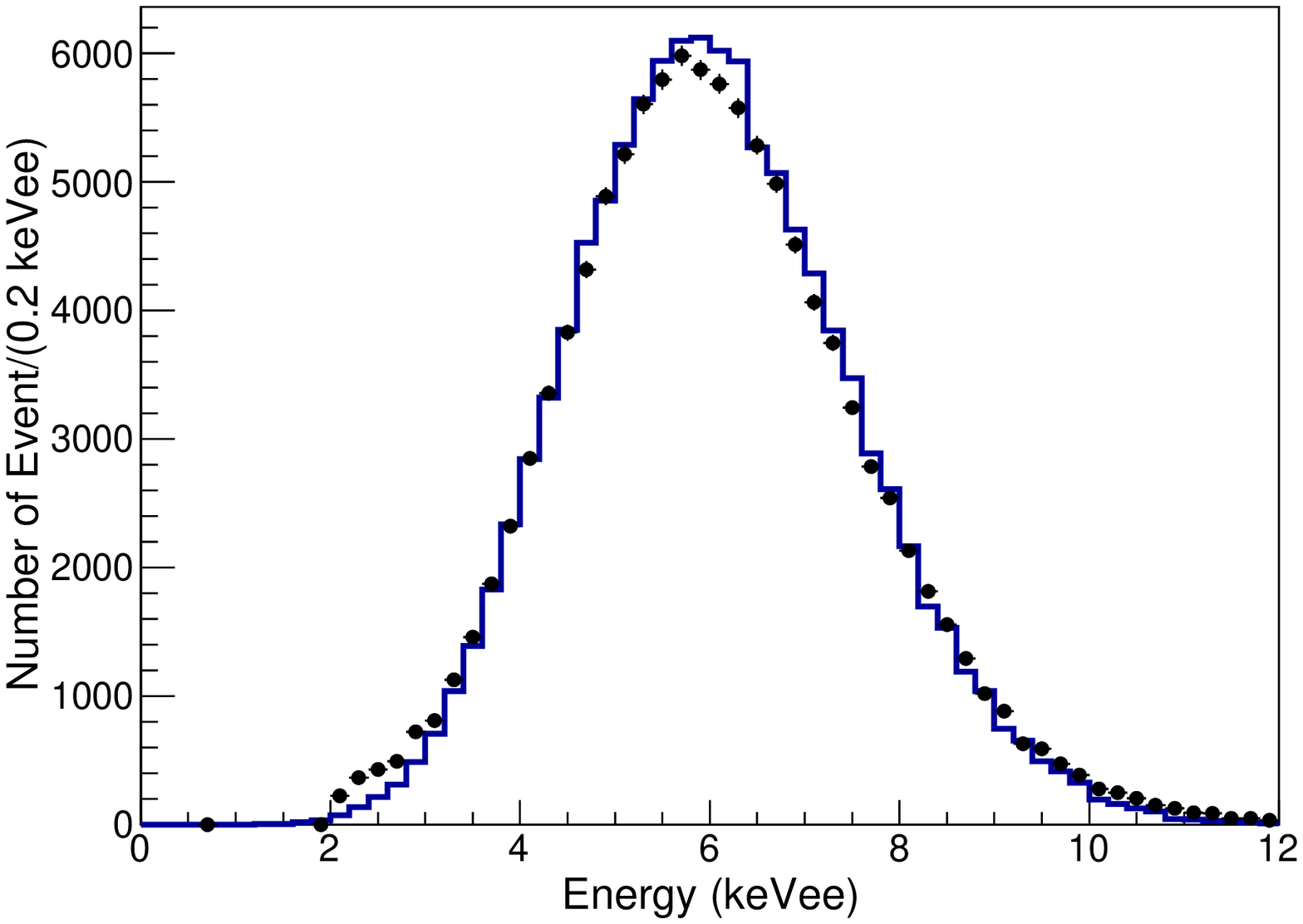}\\
(a) 59.54 keV ($^{241}$Am) & (b) 5.9 keV ($^{55}$Fe)\\
\end{tabular}
\caption[Data fit]{
(a) The quality of the simulation tuning using 59.54 keV calibration source is shown by comparing the data (filled circles) and the simulation (solid line). (b) The quality of the low-energy simulation of monoenergetic 5.9~keV photons (solid line) is compared with the 5.9~keV $^{55}$Fe calibration source data (filled circles). %Two systematic samples of the energy resolution 
}
\label{resolution_tune}
\end{cfigure1c}

\begin{figure}
\begin{center}
\includegraphics[width=\columnwidth]{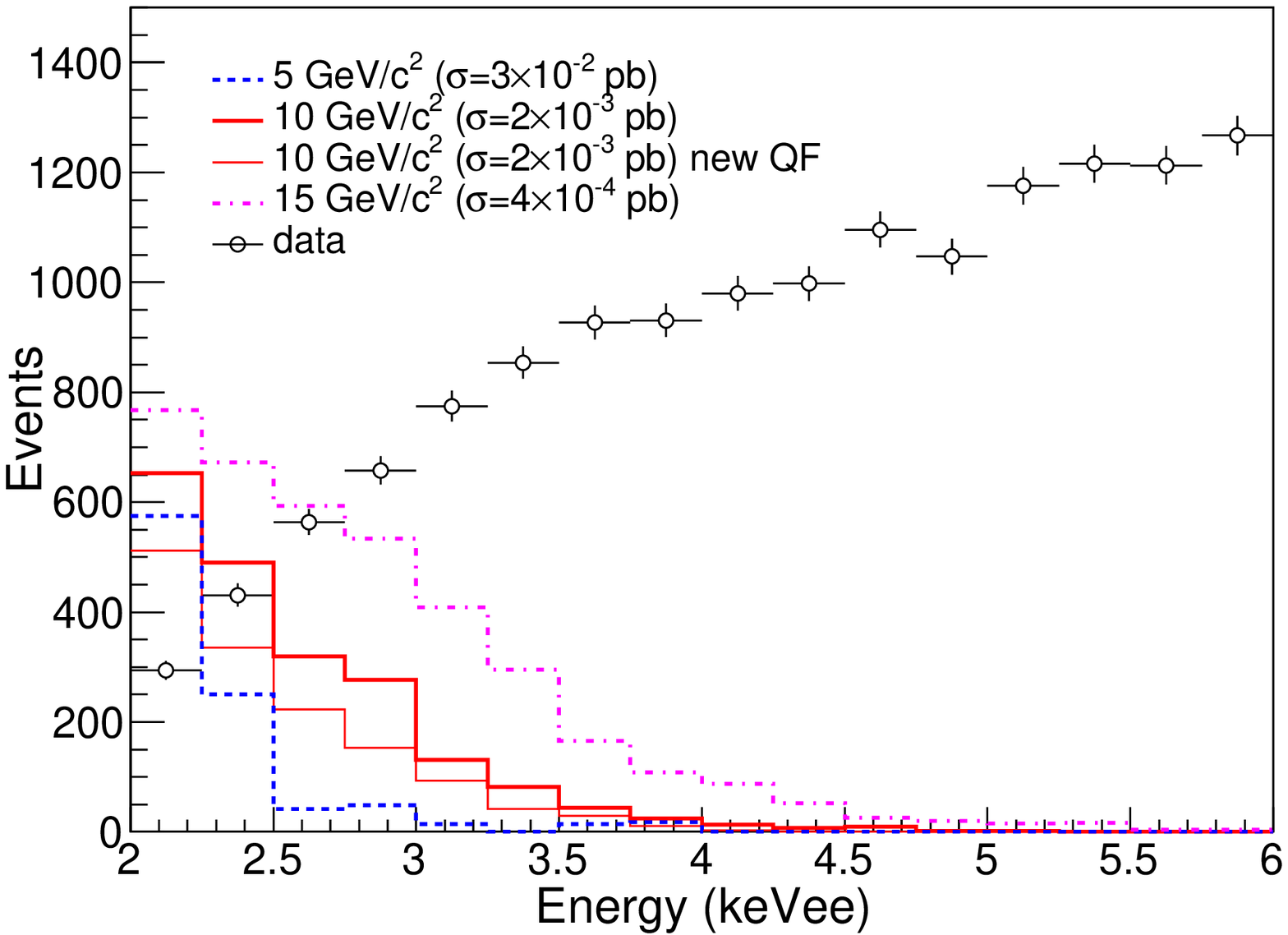}
\end{center}
\caption[BG shape]{
	Simulated WIMP energy spectra in the CsI(Tl) detector for different masses, cross sections, and new QF values shown together with the WIMP search data.
}
\label{ref:sig}
\end{figure}

We generate simulated WIMP signals based on the standard halo model~\cite{wmodel} with $v_0=220$~km/s, a galactic escape velocity of $v_{\text esc}=650$~km/s, and an average density of 0.3~GeV/cm$^{3}$ for WIMP masses between \gevcc{5} and \gevcc{20} with \gevcc{2.5} step sizes. In order to evaluate the measured nuclear recoil energy, we apply the measured quenching factors~(QF) from Ref.~\cite{kims_psd}, where QF is defined as the electron-equivalent energy divided by the nuclear recoil energy. We perform a fit of the measured quenching factor and extrapolate the results to the nuclear recoil energy below 10~keVnr, where keVnr is a nuclear recoil energy in the detector module, based on simulation. 
GEANT4-based detector simulations~\cite{geant4} are implemented for both detector responses and trigger simulations. The simulation has been tuned using 59.54~keV calibration data taken with a $^{241}$Am $\gamma$-ray source illuminating each detector module. The validity of the simulation for low energies is checked with 5.9~keV calibration data taken with a $^{55}$Fe x-ray source shown in Fig.~\ref{resolution_tune}.
We apply the selection efficiency [see Fig.~\ref{ref:fit} (a), open circles] obtained by the nuclear recoil event calibration data, which are obtained with small crystals~(3~cm$\times$3~cm$\times$3~cm) using an Am-Be neutron source~\cite{hslee_phd}, to evaluate the energy spectra of the WIMP interactions. 
Figure~\ref{ref:sig} shows the simulated WIMP energy spectra for various WIMP masses overlaid on the observed distribution after event selection. We also put the effect of new QF measurements discussed in Sec.~\ref{sec:result}.

\begin{cfigure1c}
\begin{tabular}{cc}
\includegraphics[width=0.48\textwidth]{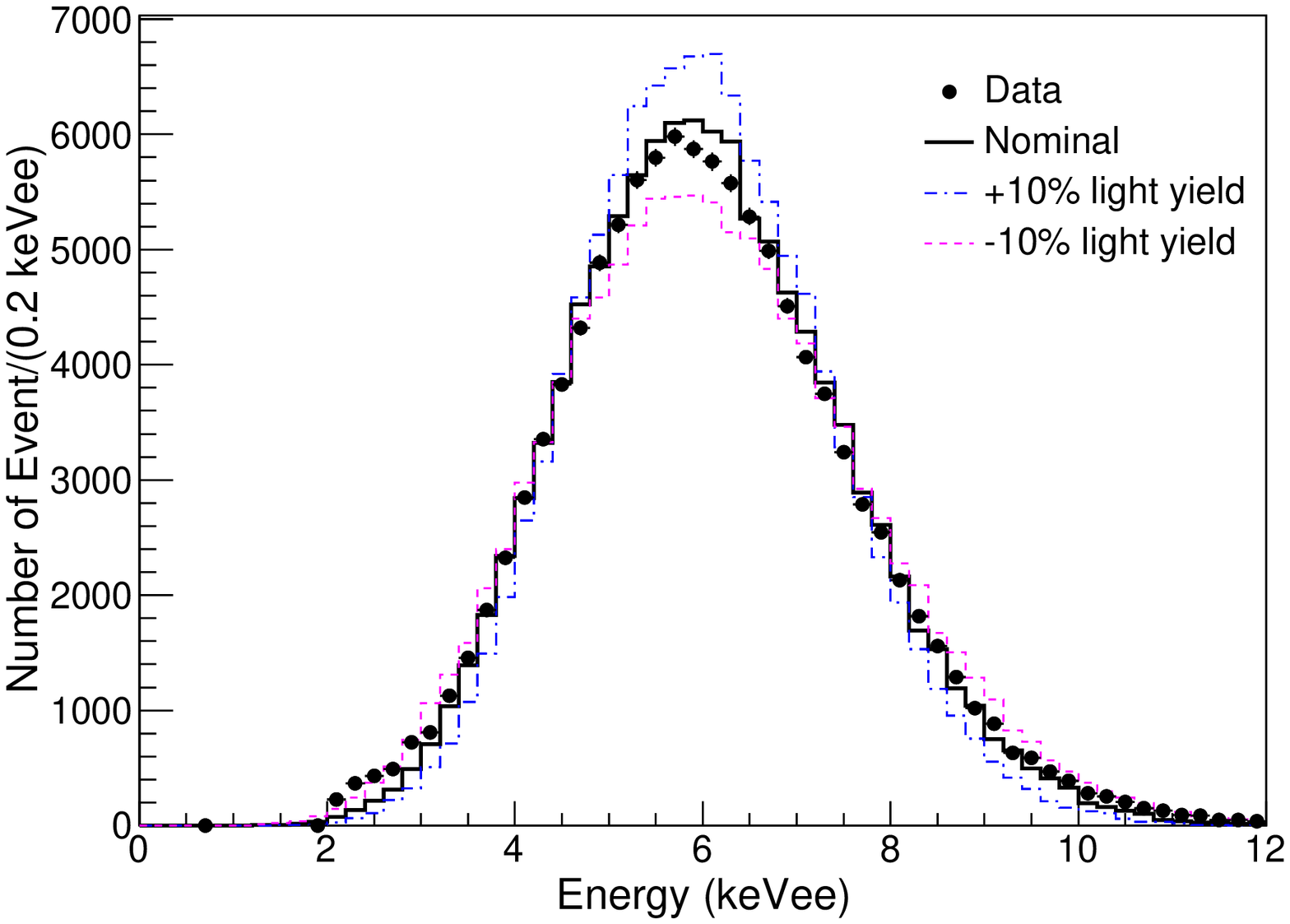}&
\includegraphics[width=0.48\textwidth]{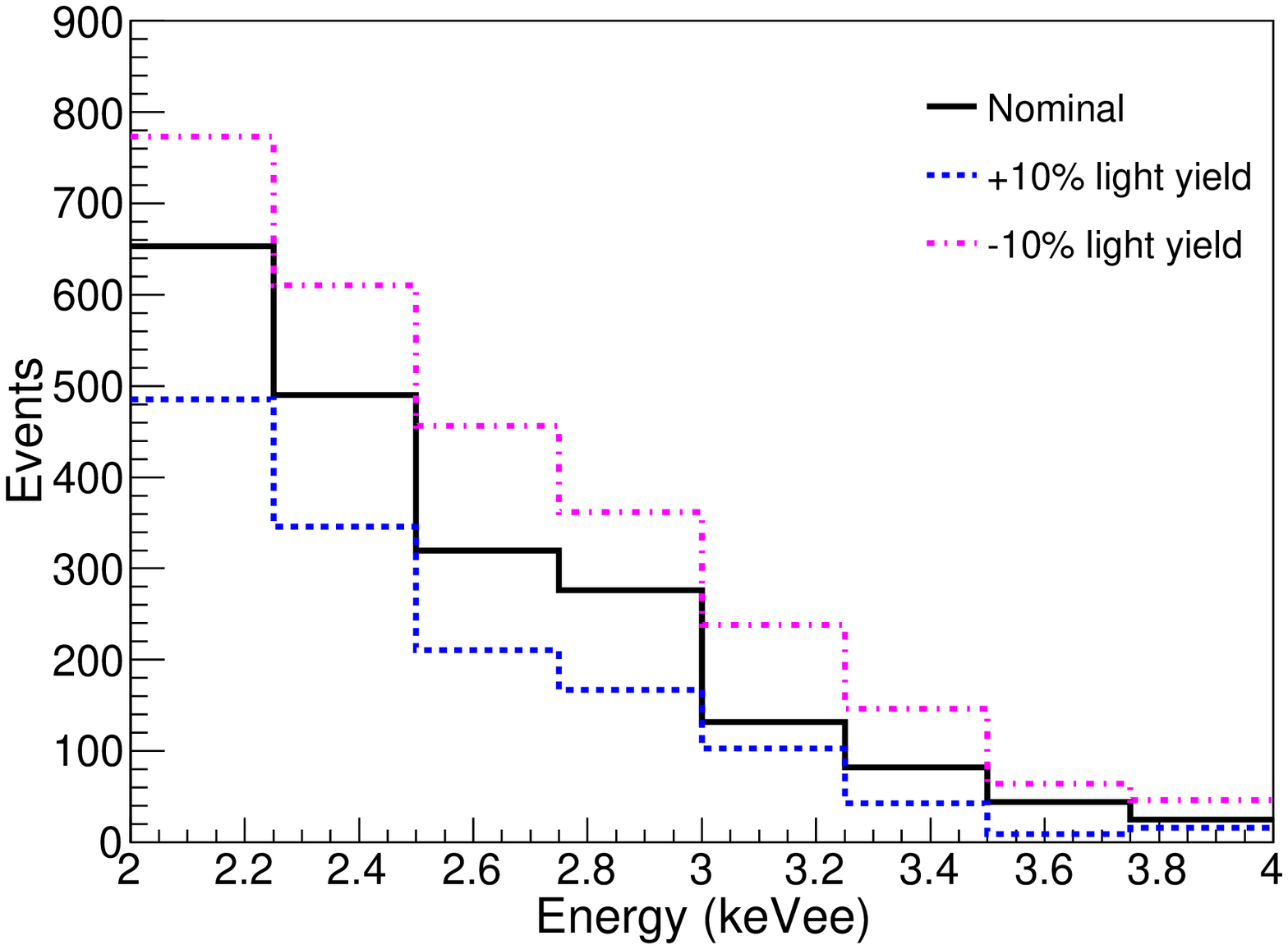}\\
(a)  & (b) \\
\end{tabular}
\caption[Data fit]{
(a) The data points show the measured energy spectrum for a 5.9~keV $^{55}$Fe calibration source in a typical detector module. The histograms are results of simulations using the light yield determined from the 59.54~keV calibration source~(solid) and yields that are 10\% higher~(dashed-dot) and lower~(dashed). 
(b) Energy spectrum of \gevcc{10} WIMP interactions in the CsI(Tl) crystal detector with the WIMP-proton spin-independent-interaction cross section $\sigma = 2\times10^{3}$~pb. 
Two models for the energy resolution determined by varying the light yield are also shown.
%Two systematic samples of the energy resolution 
}
\label{ref:resolution}
\end{cfigure1c}

%\begin{figure}
%\begin{center}
%\includegraphics[width=\columnwidth]{Eresol_det6.eps}
%\end{center}
%\caption[Data fit]{
%Energy spectrum of \gevcc{10} WIMP interactions in the CsI(Tl) crystal detector with the WIMP-proton spin-independent-interaction cross section $\sigma = 2\times10^{3}$~pb. Two systematic samples of the energy resolution are also shown by varying the light yield.
%}
%\label{ref:resolution}
%\end{figure}

\begin{table}
\begin{center}
\caption[Systematic]{
Systematic uncertainties associated with the signal processes are listed.
Assuming a \gevcc{10} WIMP interaction, we obtain the relative rate for the
twelve detector modules.
%\vspace*{1pt}
} {\footnotesize
\begin{tabular}{lccc}
\hline
Uncertainty sources    & Resolution & Calibration & Trigger \\
\hline
Relative rate (\%)    &29-35       & 12-25       & 10-20   \\
Shape change          & yes        & yes         & yes     \\
\hline
\end{tabular}}
\label{table:syst}
\end{center}
\end{table}

We consider several sources of systematic uncertainty in the extraction of the WIMP signal by propagating these uncertainties into the signal models of the measured energy distribution and expected rates.
Even though we model the energy resolution reasonably well for low energies as demonstrated in Fig.~\ref{resolution_tune}, an uncertain modeling of the energy resolution could be an important source of systematic uncertainty in the low-mass WIMP search. Because the energy resolution was extrapolated from the light output from the 59.54~keV $\gamma$-ray calibration, the influence of the light yield on the resolution of the 5.9~keV calibration data was studied.
The data points in Fig.~\ref{ref:resolution}~(a) show the measured energy spectrum from a 5.9~keV calibration source for a typical detector. The data points are well reproduced by a simulation that uses the light yield determined from the 59.54~keV $\gamma$ source, as shown by the solid histogram. Simulations that assume a 10\% higher~(lower) light yield clearly underestimate (overestimate) the width of the 5.9~keV line, as shown by the dashed-dot~(dashed) histogram in the same figure.
%Based on studies with each detector module, 
%it is found that 10\% variations of the light output can conservatively account for the 5.9~keV source resolutions for all detector modules. 
These $\pm$10\% light yield variations are then considered as conservative estimates of the systematic uncertainties associated with the energy resolution.
Figure~\ref{ref:resolution}~(b) shows the expected energy distribution for WIMP signals together with the varying light output of $+$10\% and $-$10\%.
Both the rate and the shape changes are considered as sources of systematic uncertainty.

The energy calibration of the WIMP-induced nuclear recoil signal is also an important source of systematic uncertainty.
We calibrate this energy and measure the QF in a separate test using nuclear recoils produced by elastic scattering of 2.63~MeV neutrons of $^{3}$H(p,n)$^3$He reactions with 3.4~MeV proton beam for the energy greater than 10~keVnr~\cite{kims_psd}.
Extrapolation into the recoil energy region below 10~keVnr was done with a fit of the measured QF with the consistency checked by a SRIM-based simulation~\cite{srim} as described in Ref.~\cite{quench_kims}. We assign a 15-\% systematic uncertainy on the QF based on the results of the fit. 
We consider the rate and shape changes from QF variations as systematic uncertainties. 
%We perform a fit of the measured quenching factor and extrapolate the results to the nuclear recoil energy below 10keVnr. 
%However the measurement have been performed the recoil energy greater than 10~keVnr.
%From the measured QF uncertainty, we determine the systematic errors associated with rate and shape changes.

The trigger efficiency modeling also contributes to the systematic uncertainty.
We simulate the trigger requirement of two reconstructed PEs within 2~$\mu$s in each of the module's two PMTs. To account for possible uncertainty in the trigger simulation, we also employed a data-driven technique in which we assume a flat energy distribution for multiple-module events that are mostly due to the Compton scattering of high energy $\gamma$ backgrounds. We then estimate the trigger efficiency of the measured data.
The difference between the trigger simulation and the data-driven technique is treated as a systematic uncertainty.
Table~\ref{table:syst} shows a summary of the systematic uncertainties for the case of \gevcc{10} WIMP signals.

\section{Results and Discussion}
\label{sec:result}

\begin{figure}
\begin{center}
\includegraphics[width=\columnwidth]{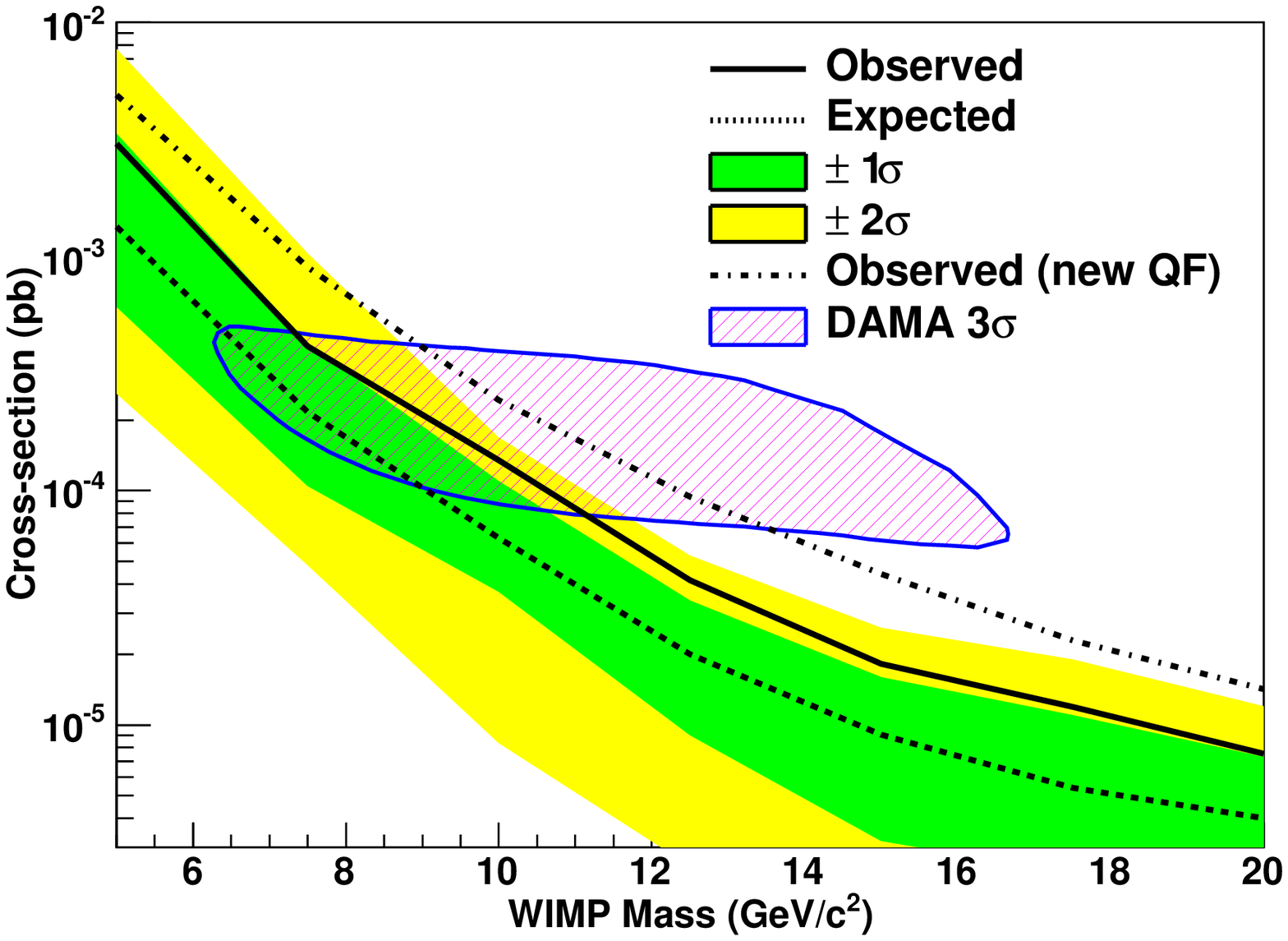}
\end{center}
\caption[Limit]{
Observed (solid line) and median-expected (dotted line) 90\% confidence level exclusion limits on the WIMP-nucleon spin-independent cross section (assuming the background-only hypothesis) are shown together with the low-mass WIMP-induced DAMA $3\sigma$ allowed signal region. The dark and light-shaded bands indicate the one and two standard deviation probability regions in which the limits are expected to fluctuate in the absence of a signal. The dotted-dashed line~(new QF) indicates the observed limit using the QF from the MARLOWE-based simulation~\cite{quench_kims,quench_kims1}.
}
\label{ref:limit}
\end{figure}

\begin{table*}
\begin{center}
\caption[Limit]{
The median-expected (Exp) 90\% C.L. upper limits assuming the background-only hypothesis and  QF obtained from a neutron beam test are shown with the corresponding observed (Obs) limits on the WIMP-nucleon spin-independent cross sections for seven WIMP mass hypotheses within the $\gevcc{5\le m_{W}\le20}$ range. The limits with the new QF obtained from the MARLOWE-based simulation~\cite{quench_kims,quench_kims1} are also shown.
%\vspace*{1pt}
} {\footnotesize
%\begin{ruledtabular}
\begin{tabular}{cccccccc}
\hline
WIMP mass (\gevccnoarg)   &5 & 7.5 & 10 & 12.5 & 15 & 17.5 & 20 \\
\hline
Exp (pb)                  & 1.3$\times 10^{-3}$& 2.2$\times 10^{-4}$ & 6.3$\times 10^{-5}$ & 2.0$\times 10^{-5}$ & 9.1$\times 10^{-6}$ & 5.4$\times 10^{-6}$ & 4.0$\times 10^{-6}$ \\
Obs (pb)                  & 3.0$\times 10^{-3}$& 4.1$\times 10^{-4}$ & 1.3$\times 10^{-4}$ & 4.2$\times 10^{-5}$ & 1.8$\times 10^{-5}$ & 1.2$\times 10^{-5}$ & 7.5$\times 10^{-6}$ \\
\hline
Exp/new QF (pb)           & 1.8$\times 10^{-3}$& 4.2$\times 10^{-4}$ & 1.1$\times 10^{-4}$ & 4.4$\times 10^{-5}$ & 1.9$\times 10^{-5}$ & 1.0$\times 10^{-5}$ & 7.2$\times 10^{-6}$ \\
Obs/new QF (pb)           & 4.9$\times 10^{-3}$& 9.0$\times 10^{-4}$ & 2.4$\times 10^{-4}$ & 9.4$\times 10^{-5}$ & 4.4$\times 10^{-5}$ & 2.3$\times 10^{-5}$ & 1.4$\times 10^{-5}$ \\
\hline
\end{tabular}}
\label{table:limit}
\end{center}
\end{table*}

We extract upper limits on the WIMP-proton spin-independent cross section using a Bayesian likelihood~\cite{pdg} formed as a product of likelihoods over bins of the measured energy distribution between 2~keVee and 4~keVee for all detector modules.
We assume uniform priors for the signal and background rates and Gaussian priors for each systematic uncertainty. We further assume that there are no negative signals.
We consider the possibility of correlated rate and shape uncertainties as well as the uncorrelated bin-by-bin statistical uncertainties.
A 90\% confidence level~(C.L.) limit is determined such that 90\% of the posterior density of the WIMP-nucleon cross section falls below the limit.
The expected 90\% C.L. limits are calculated based on the expected backgrounds from 2,000 simulated experiments.
The observed 90\% C.L. limits are calculated from the data.
The obtained median limits are listed in Table~\ref{table:limit}. Over the range of all WIMP masses, we find an overall excess of the observed limits that is about one standard deviation over the median-expected limit as shown in Fig.~\ref{ref:limit}. Therefore, we conclude that our results are consistent with null signals for WIMP interactions.
In Fig.~\ref{ref:limit}, we also include three standard deviation contours associated with the low-mass WIMP interpretation of the DAMA annual modulation signal. As one can see from this figure, we cover most, but not all, of this DAMA signal region. These are the first results that specifically address, using a similar crystal detector, the low-mass WIMP allowed region from the DAMA experiment.

Because of possible issues on the QF measurements with the scintillating detectors~\cite{quenching}, we have also studied the responses of nuclear recoil events in the CsI(Tl) detector using $^2$H($^2$H,n)$^3$He reactions from a commercial neutron generator~\cite{quench_kims}. From an analysis discussed in Ref.~\cite{quenching,quenching1}, we find that the measured spectra of the nuclear recoil events in the CsI(Tl) crystal are well reproduced by the MARLOWE program~\cite{marlowe} used in conjuction with a Geant4-based detector simulation~\cite{quench_kims1}. The details of the simulation as well as preliminary QF results using the MARLOWE program can be found in Ref.~\cite{quench_kims}.
Because the QF from this simulation is approximately 30\% lower than the results of the previous measurements~\cite{kims_psd,csi_quench,csi_quench1} and the SRIM-based simulation, we evaluate the expected and observed limits of the WIMP-proton cross section using this new QF in Table~\ref{table:limit}.
We also include the observed limit with the new QF in the limit plot of Fig.~\ref{ref:limit}. In this case, the coverage of the DAMA signal region is narrower. However, here the contour of DAMA signal region was calculated with old QF measurement of NaI(Tl) crystals~\cite{savage}. If we interpret the DAMA signal with newly measured QF considering the issues of the QF measurement, which brought on  lowered QF of NaI(Tl) crystals, the allowed DAMA region also move to same direction of the observed limit as discussed in Ref.~\cite{quenching}.  %but we have not included effects of the lower QF reported for the DAMA NaI(Tl) crystals in Ref.~\cite{quenching} on the contours of the allowed DAMA region.

\section{Conclusion}
This Letter presents the results of a search for low-mass WIMPs using CsI(Tl) crystal detectors. We use a reduced the analysis threshold and search for WIMPs with masses below \gevcc{20}. We find no significant evidence of a WIMP signature in our data and set 90\% C.L. upper limits that partially cover the DAMA allowed signal region for low-mass WIMPs assuming the standard halo model.  We are now replacing the PMTs with higher light output types that will reduce the trigger and analysis threshold. Once the upgrade is complete, we expect the experimental sensitivity will cover the DAMA allowed $3\sigma$ signal region but also  parameter values favored by other experiments that report hints of low-mass WIMPs~\cite{CRESST,CoGent,CDMS-low}.

\begin{acknowledgments}
We are very grateful to the Korea Midland Power Co. and Korea Hydro and Nuclear Power Co. for providing the underground laboratory space at Yangyang. This research was supported by the research center program of the Institute for Basic Science~(IBS) in Korea, the WCU program (R32-10155), and the National Research Foundation of Korea (NRF-2011-0031280 and NRF-2011-35B-C00007). H.S.L. was supported by the Ewha Womans University Research Grant of 2012. 
\end{acknowledgments}

%\section*{References}

\end{document}